\documentclass[epjST]{svjour}
\usepackage{color}
\usepackage{graphics}
\usepackage{epsfig}
\usepackage[latin1]{inputenc}
\usepackage{amssymb}
\usepackage{amsmath}

\graphicspath{{./}{./Figures/}}

\usepackage{latexsym}
\usepackage{mathrsfs}
\usepackage{bm}
\usepackage{comment}

\newcommand{\m}[1]{\begin{pmatrix}#1\end{pmatrix}}

\newcommand{\matr}[1]{{{\bm{#1}}}}    
\renewcommand{\vec}[1]{{\bm{#1}}}    

\begin{document}
\title{Delay-induced chimeras in neural networks with fractal topology}
\author{Jakub Sawicki\inst{1}\fnmsep\thanks{\email{zergon@gmx.net}} \and Iryna Omelchenko\inst{1} \and Anna Zakharova\inst{1} \and Eckehard Sch\"{o}ll\inst{1}}
\institute{Institut f\"{u}r Theoretische Physik, Technische Universit\"{a}t Berlin, Hardenbergstr. 36, 10623 Berlin, Germany}
\abstract{
We study chimera states, which are partial synchronization patterns consisting of spatially coexisting domains of coherent (synchronized) and incoherent (desynchronized) dynamics, in ring networks of FitzHugh-Nagumo oscillators with fractal connectivities. In particular, we focus on the interplay of time delay in the coupling term and the network topology. In the parameter plane of coupling strength and delay time we find tongue-like regions of existence of chimera states alternating with regions of coherent dynamics. We show analytically and numerically that the period of the synchronized dynamics as a function of delay is characterized by a sequence of piecewise linear branches.
In between these branches various chimera states and other partial synchronization patterns are induced by the time delay. By varying the time delay one can deliberately choose and stabilize desired spatio-temporal patterns. 
} 
\maketitle
\section{Introduction}
Synchronization of coupled nonlinear oscillators is a widely studied field of nonlinear dynamics, which has a plethora of applications in natural and and technological systems~\cite{PIK01,BOC06a,STE17}. Recent interest has focussed on partial synchronization patterns like chimera states, which consist of spatially coexisting domains of coherent (synchronized) and incoherent (desynchronized) dynamics~\cite{PAN15,SCH16b}. They were first found theoretically in systems of phase oscillators~\cite{KUR02a,ABR04}, and later also in a large variety of different systems including time-discrete maps~\cite{OME11,SEM15a,VAD16,BOG16,BUK17,SEM17}, time-continuous chaotic models~\cite{OME12}, neural systems~\cite{OME13,HIZ13,VUE14a,OME15,TSI16,TSI17,CHO18}, Boolean networks~\cite{ROS14}, population dynamics~\cite{BUS15,HIZ15,BAN16}, quantum oscillators~\cite{BAS15}, and in higher spatial dimensions~\cite{OME12a,PAN15,SHI04,MAI15,XIE15a,TOT17,BRU18}. Alternating ~\cite{HAU15} as well as amplitude-mediated~\cite{SET13,SET14}, and pure amplitude chimera and chimera death states~\cite{ZAK14,ZAK15b,BAN15} were discovered, and a universal classification scheme has recently been introduced~\cite{KEM16}.
Chimera states have been associated with real-world phenomena like uni-hemispheric sleep~\cite{RAT00,RAT16}, bump states in neural systems~\cite{LAI01,SAK06a}, epileptic seizures~\cite{ROT14,AND16}, power grid failure~\cite{MOT13a}, or collective dynamics in social systems~\cite{GON14}. 
Experimentally, chimeras have been found in optical~\cite{HAG12,BRU18}, chemical~\cite{TIN12,NKO13,TOT17} systems, mechanical~\cite{MAR13,KAP14}, electronic~\cite{LAR13,GAM14}, optoelectronic delayed-feedback~\cite{LAR15} and electrochemical~\cite{WIC13,SCH14a} oscillator systems, Boolean networks~\cite{ROS14}, and optical combs~\cite{VIK14}. Chimera states have also been shown to be robust against inhomogeneities of the local dynamics and coupling topology~\cite{OME15}, against noise~\cite{LOO16}, or they might even be induced by noise~\cite{SEM15b,SEM16,ZAK17}. 

The topology of the network has been found to play a crucial role in inducing chimera states. While earlier work has focussed on simple nonlocal coupling schemes like rings or two-module structures, chimeras have also been found in all-to-all coupled networks~\cite{SET14,YEL14,BOE15,SCH15a,SCH15e}, as well as in more complex coupling topologies. Of particular interest are networks with hierarchical connectivities, arising in neuroscience as shown by Diffusion Tensor Magnetic Resonance Imaging analysis, which found a hierarchical (quasi-fractal) connectivity of the neuron axons network~\cite{KAT09,EXP11,KAT12,KAT12a,PRO12}. Such a network topology can be realized using a Cantor algorithm starting from a chosen base pattern~\cite{OME15,ULO16,BON18,CHO18}, and is in the focus of our present study. 

Control of chimera states by extending their lifetime and fixing their spatial position is an important issue~\cite{SIE14c,BIC15,OME16,OME18}. A well-known method for stabilization or destabilization of complex patterns in networks is time delay~\cite{AHL04,FLU13,SAH17}. Time-delayed feedback or coupling has been shown to be a versatile method for controlling chimera states \cite{GJU17,ZAK17a,SAW17}. The goal of this paper is to study the influence of time delay on chimera states in networks of FitzHugh-Nagumo oscillators with fractal connectivity, and to demonstrate how by varying the time delay one can stabilize chimera states in the network.

\section{The Model}
\label{sec:model}
The FitzHugh-Nagumo (FHN) is a paradigmatic model for neural systems \cite{FIT61,NAG62}, but is also used to describe chemical \cite{SHI04} and optoelectronic \cite{ROS11a} oscillators and nonlinear electronic circuits \cite{HEI10}. We consider a ring of $N$ identical FHN oscillators with fractal coupling topology, which is given by the adjacency matrix $\matr{G}$ with a circulant structure. The dynamical equations for the variable $\vec{x}_k=(u_k, v_k)^T \in \mathbb{R}^2$, where $u_k$ and $v_k$ are the activator and inhibitor variables, respectively, are:
\begin{align}
\vec{\dot{x}}_i(t) &=  \vec{F}(\vec{x}_i(t)) +  \frac{\sigma}{g} \sum^{N}_{j=1}G_{ij} \matr{H}[\vec{x}_j(t-\tau)-\vec{x}_i(t)]
\label{eqn:gen1}
\end{align}
with $i \in \{1,...,N\}$ modulo $N$, and the delay time $\tau$. The dynamics of each individual oscillator is governed by 

\begin{eqnarray}
\label{eq:localdyn}
\vec{F}(\vec{x})=
\left(\!
\begin{array}{*{1}{c}}
\varepsilon^{-1}(u-\frac{u^3}{3}-v)\\
u + a
\end{array}
\!\right),
\end{eqnarray}
where $\varepsilon > 0$ is a small parameter characterizing a time scale separation, which we fix at $\varepsilon = 0.05$ throughout the paper. Depending on the threshold parameter $a$ the FHN oscillator exhibits either oscillatory ($|a|<1$) or excitable ($|a|>1$) behavior. We consider the oscillatory regime ($a=0.5$) in this work. The parameter~$\sigma$ denotes the coupling strength,  and $g=\sum^{N}_{j=1}G_{ij}$ is the number of links for each node (corresponding to the row sum of $\matr{G}$). The interaction is realized through diffusive coupling with coupling matrix 
\begin{eqnarray}
\matr{H}=\m{\varepsilon^{-1}\cos \phi& \varepsilon^{-1} \sin \phi\\ -\sin \phi&\cos \phi}
\end{eqnarray}
In accordance with Omelchenko et al.~\cite{OME13}, throughout the manuscript we fix the coupling phase $\phi = \frac{\pi}{2}-0.1$.

\subsection{Fractal topology}
Fractal topologies can be generated using the Cantor construction algorithm for a fractal set~\cite{MAN83,FED88}. This iterative hierarchical procedure starts from a \emph{base pattern} or initiation string $b_{init}$ of length $b$, where each element represents either a link ('$1$') or a gap ('$0$'). The number of links contained in $b_{init}$ is referred to as $c_1$. In each iterative step, each link is replaced by the initial base pattern, while each gap is replaced by $b$ gaps. Thus, each iteration increases the size of the final bit pattern, such that after $n$ iterations the total length is $N=b^n$. We call the resulting connectivity fractal or hierarchical. 
Using the resulting string as the first row of the adjacency matrix $\matr{G}$, and constructing a circulant adjacency matrix $\matr{G}$ by applying this string to each element of the ring, a ring network of $N=b^n$ nodes with hierarchical connectivity is generated ~\cite{OME15,TSI16,HIZ15}. Here we slightly modify this procedure by including an additional zero in the first instance of the sequence~\cite{ULO16}, which corresponds to the delayed self-coupling. Therefore, there is no net effect of the diagonal elements of the adjacency matrix $G_{ii}$ on the network dynamics. Without our modification, this would lead to a breaking of the base pattern symmetry, i.e., if the base pattern is symmetric, the resulting coupling topology would not be so, since the first link to the right is missing from the final link pattern.
Our procedure, in contrast, ensures the preservation of an initial symmetry of $b_{init}$ in the final link pattern, which is crucial for the observation of chimera states, since asymmetric coupling leads to a drift of the chimera ~\cite{BIC15,OME16}. Thus, a ring network of $N=b^n+1$ nodes is generated.

\subsection{Chimera states}

In the following, we consider the network generated with base pattern $b_{init}=(11011)$ after four iterative steps, resulting in a ring network of $N=5^4 + 1= 626$ nodes. Our choice is motivated on the one hand by previous studies of chimera states in nonlocally coupled networks~\cite{OME13,OME15a}, where it has been shown that an intermediate coupling range is crucial for the observation of chimera states, too large and too small numbers of connections make this impossible. On the other hand, it has been demonstrated that hierarchical networks with higher clustering coefficient promote chimera states~\cite{OME15,ULO16}. For the fractal topology considered here the clustering coefficient $C$ introduced by Watts and Strogatz~\cite{WAT98} is calculated as $C=0.428$. In our fractal network we obtain an effective coupling radius $\bar{r}=\frac{c_1^n}{2N}=0.2$, namely half the link density, as derived in Ref.\cite{ULO16}, which is much smaller than the coupling radius $r$ for which chimeras have been observed in regular nonlocally coupled networks~\cite{OME13,OME15a}. 

\begin{figure}[htbp]
\centering
\includegraphics[width = 1.15\textwidth]{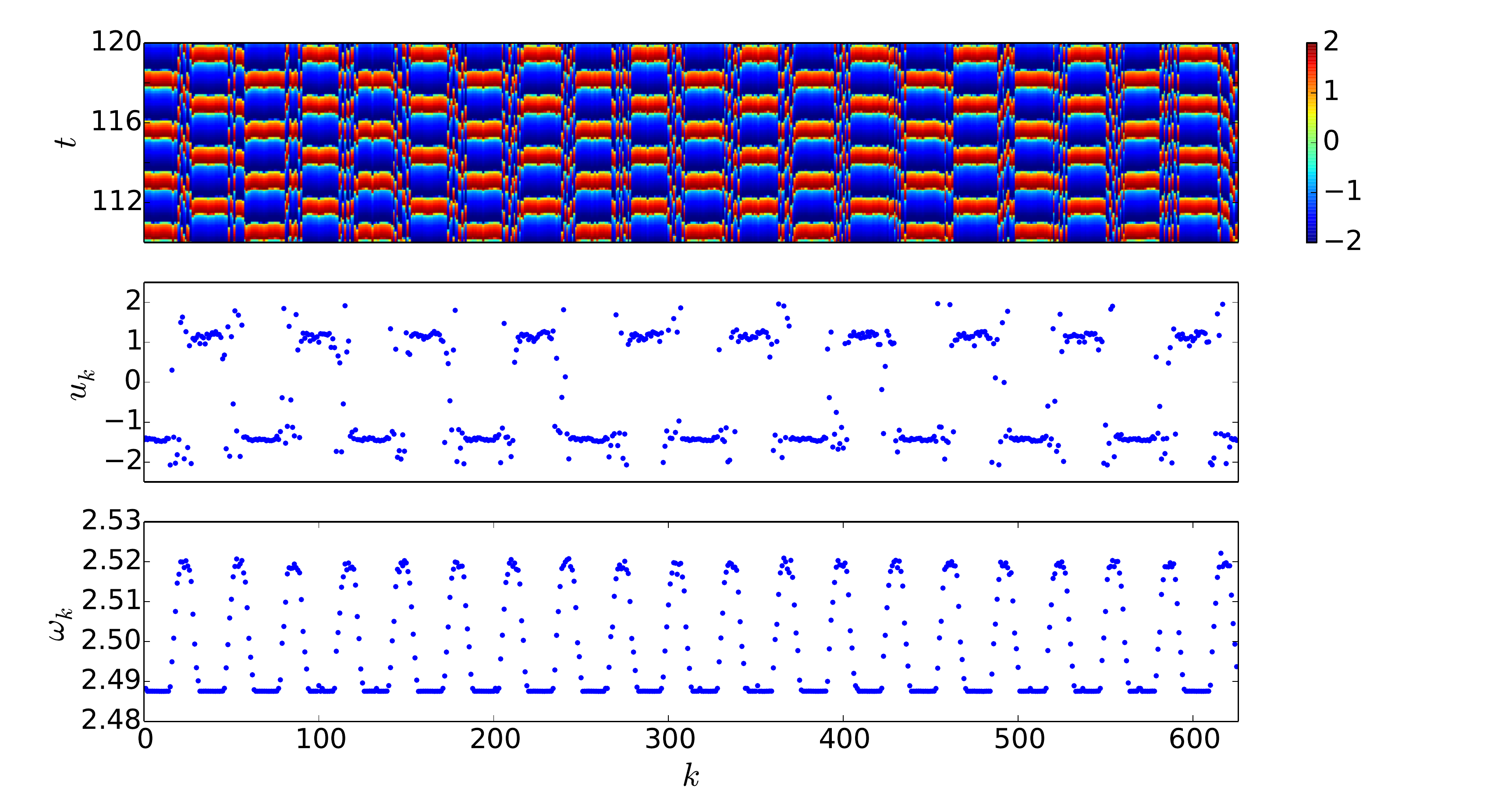}
\caption{(Color online) Chimera state in the case $\tau = 3.6$ and $\sigma=0.05$ for $b_{init}=(11011)$, $n=4$, $N=626$, $a=0.5$, $\varepsilon=0.05$, and $\phi = \frac{\pi}{2}-0.1$. Random initial conditions were used. The three panels correspond to the same simulation: Space-time plot of $u_k$ (upper panel), snapshot of variable $u_k$ at $t=50000$ (middle panel), and mean phase velocity profile $\omega_k$ (bottom panel).  
}
\label{fig:1}
\end{figure}

\section{Influence of time delay}

Figure\,\ref{fig:1} demonstrates a chimera state in the system~(\ref{eqn:gen1}) for time delay~$\tau=3.6$, obtained numerically for random initial conditions. We analyze the space-time plot (upper panel), the final snapshot of the activator variables $u_k$ at $t=50000$ (middle panel), and the phase velocities $\omega_k$ of the oscillators (bottom panel). The mean phase velocities of the oscillators are calculated as $\omega_k = 2\pi S_k / \Delta T$, $k=1,...,N,$ where $S_k$ denotes the number of complete rotations realized by the $k$th oscillator during the time $\Delta T$. Throughout the paper we used $\Delta T = 10000$. Oscillators from coherent domains are phase-locked and have equal mean frequencies. Arc-like profiles of the mean phase velocities for oscillators from the incoherent domain are typical for chimera states.     

To uncover the influence of time delay introduced in the coupling term in system~(\ref{eqn:gen1}), we analyze numerically the parameter plane of coupling strength~$\sigma$ and delay time~$\tau$. Fixing the network parameters $b_{init}=(11011)$, $n=4$, $N=626$, $a=0.5$, and $\varepsilon=0.05$, we choose random initial conditions, and vary the values of $\sigma$ and $\tau$.

\begin{figure}[htbp]
\centering
\includegraphics[width = 1.03\textwidth]{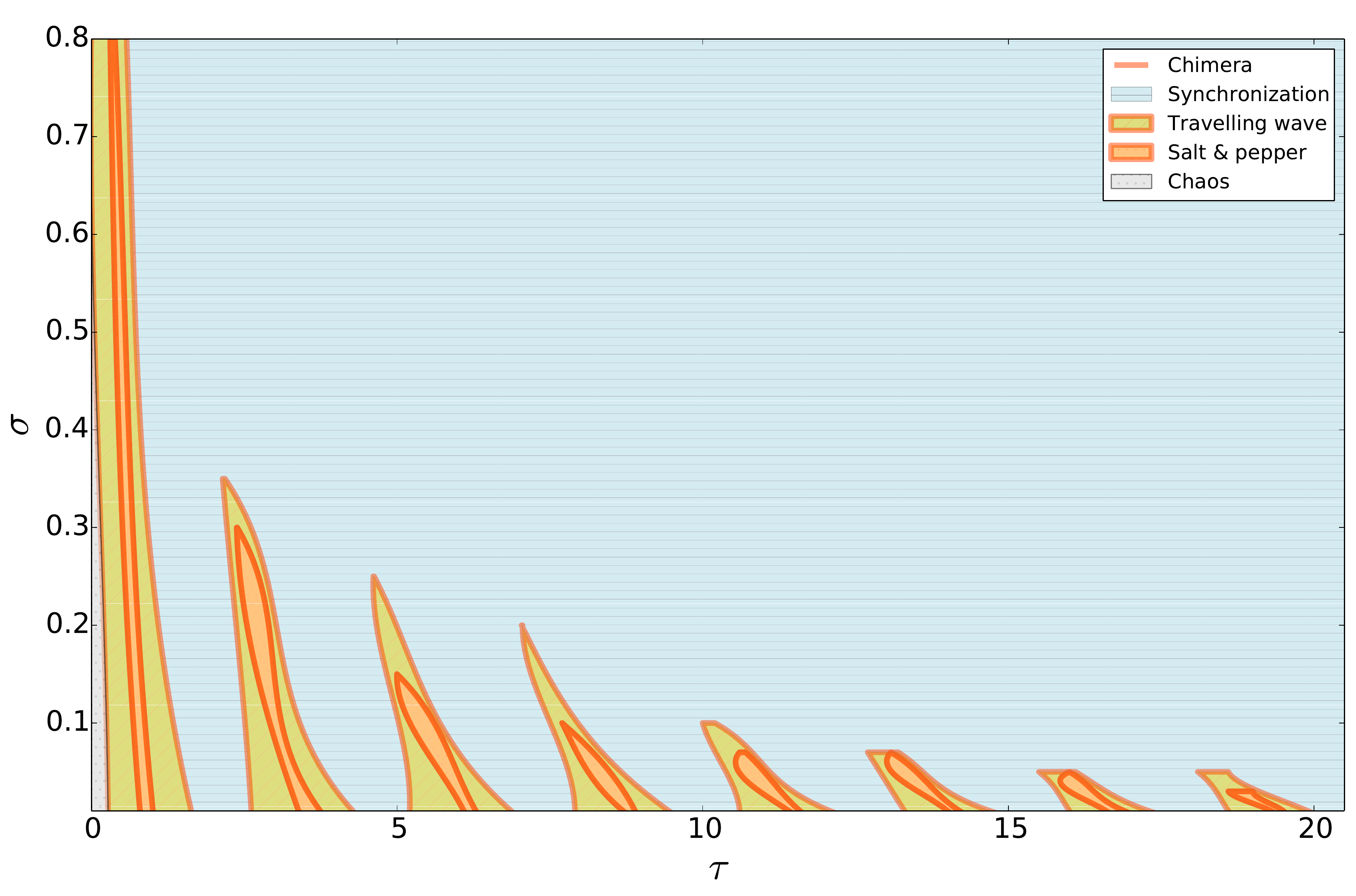}
\caption{(Color online) Chimera tongues: Chimeras occur on the boundaries (red curves) between in-phase synchronization (horizontally striped blue region), coherent traveling waves (diagonally striped yellow-green regions), and ``salt and pepper" dynamics (dotted red regions) in the parameter plane~$(\tau,\sigma)$. Below the first traveling wave region we can observe chaos (dotted grey region at small $\tau,\sigma$). Random initial conditions were used for all numerical simulations. Other parameters as in Fig.\,\ref{fig:1}.
}
\label{fig:2}
\end{figure}

Figure\,\ref{fig:2} shows the map of regimes in the parameter plane~$(\tau,\sigma)$. In the undelayed case $\tau=0$ we cannot observe chimera states for random initial conditions. The introduction of small time delay for weak coupling strength does not change the behavior and the system stays in the completely incoherent regime characterized by chaotic dynamics (grey dotted region). Nevertheless, for larger values of coupling strength~$\sigma$ chimera states can be observed for small $\tau$. With increasing delay $\tau$ we observe a sequence of tongue-like regions, which are bounded by red curves, on which chimera states occur. These regions appear in between larger areas of coherent structures: fully synchronized states (blue regions with horizontal stripes) alternating with coherent traveling waves, where all nodes oscillate with the same phase velocity (yellow-green regions with diagonal stripes). Inside the tongues we can observe ``salt and pepper" states, which are characterized by strong variations on very short length scales, so that the dynamical patterns have arbitrarily short wavelengths~\cite{BAC14,SEM17} (red dotted regions). Closer inspection of the chimera tongues shows that increasing $\tau$ reduces the size of the tongues, and also decreases the maximum values of $\sigma$ for which chimera states are observed. Moreover, one can easily see that chimera regions appear at $\tau$ values close to half-integer multiples of the period of the uncoupled system $T \approx 2.3$. 

\begin{figure}[htbp]
\centering
\includegraphics[width = 1.0\textwidth]{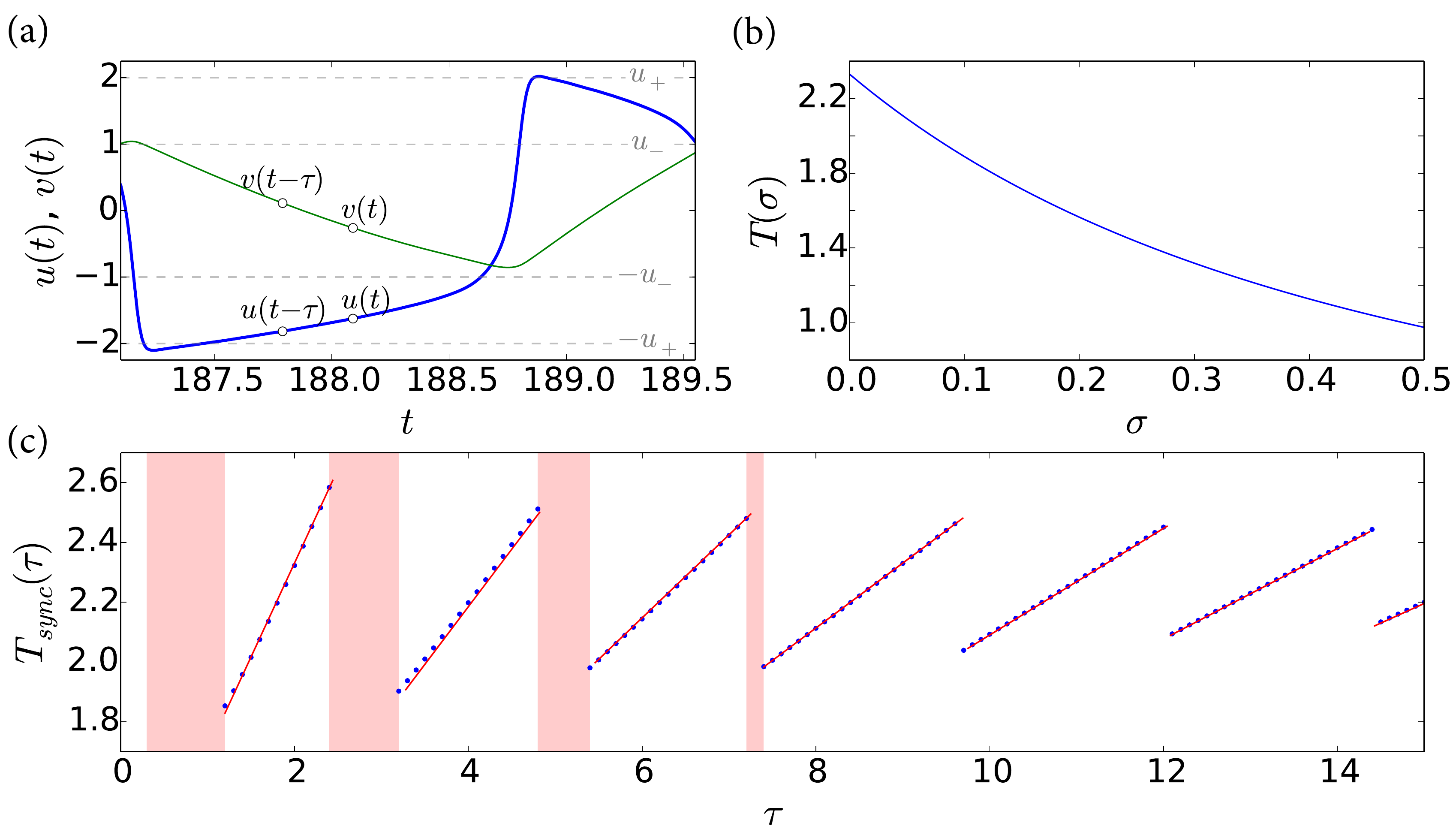}
\caption{(Color online) Analytical approximation of the period $T$ for an FHN system with delayed coupling: (a) Limit cycle of the variables $u(t)$ (dark blue line) and $v(t)$ (light green line) of a single FHN oscillator with delayed feedback representing the synchronized state of Eq.\,\eqref{eqn:gen1} for $\tau = 2.1$ and $\sigma = 0.15$. The dashed grey lines indicate $\pm u_{\pm}$ respectively, given by $u_{+}=2$ and $u_{-}=1$. (b) Period $T$ vs. $\sigma$ of the FHN oscillator given by Eq.\,\eqref{eq:T_sigma}, valid for parameters from the incoherent regimes in Fig.\,\ref{fig:2}. As the proportionality factor we assume $1+e(\varepsilon)$, with $e(0.05)=0.3$. (c) Period of the synchronized solution $T_{sync}$ vs. delay time $\tau$. Comparison of numerics (dots) and analytics (lines), given by Eq.\,\eqref{eq:T_tau_gen} for $\sigma = 0.15$. The red shaded regions correspond to the tongues in Fig.\,\ref{fig:2}. Other parameters for all panels as in Fig.\,\ref{fig:1}.}
\label{fig:3}
\end{figure}

In many delay systems one expects resonance effects if the delay is an integer or half-integer multiple of the period of the uncoupled system~\cite{HOE05,YAN06}. The undelayed part of the coupling term in Eq.~(\ref{eqn:gen1}) is the most important part in case of incoherence (see tongues in Fig.\,\ref{fig:2}) and can be rewritten as follows, neglecting $\cos \phi \ll 1$ and setting $\sin \phi \approx 1$ (it is possible to keep $\phi$, but it complicates the algebra, see Eq.\,\eqref{eq:T_phaselag}):
\begin{equation}
\begin{aligned}
\label{eq:incoh}
\varepsilon \dot{u} &= u - \frac{u^3}{3} - (1+\sigma) v\\
\dot{v} &= (1+\sigma) u + a
\end{aligned}
\end{equation}
Similar to Brandstetter~\cite{BRA09} we employ an analytic approximation for the period of the oscillation defined by Eq.\,\eqref{eq:incoh}. We consider slow motion on the falling branches of the $u$-nullcline given by $(1+\sigma) v = u - \frac{u^3}{3}$ and hence $(1+\sigma) \dot{v} = \dot{u}(1 - u^2)$, which gives:
\begin{eqnarray}
\dot{u}=\frac{(1+\sigma)^2u+(1+\sigma)a}{1-u^2}
\end{eqnarray}
It is possible to integrate this equation analytically from $\pm u_{+}$ to $\pm u_{-}$, which are approximately the limits of the slow parts of the $u$-nullcline (see Fig.\,\ref{fig:3}a), given by $u_{+}= 2$ and $u_{-}=1$. With this we obtain a rough approximation of the intrinsic period $T(\sigma)$ of the coupled system, neglecting the fast parts of the trajectory $u(t)$:
\begin{eqnarray}
T(\sigma) \propto (1+\sigma)^{-2}\left[u_+^2-u_-^2+\left(1-\left(\frac{a}{1+\sigma}\right)^2 \right)\ln \frac{a^2-(1+\sigma)^2u_-^2}{a^2-(1+\sigma)^2u_+^2}\right]
\label{eq:T_sigma}
\end{eqnarray}
As we can see in Fig.\,\ref{fig:3}(b) the period $T$ decreases with increasing $\sigma$. Therefore, due to the resonance condition of $\tau$ with respect to the intrinsic period $T$, the chimera tongues are shifted to the left with increasing coupling strength $\sigma$. 

In the case of complete synchronization (blue region in Fig.\,\ref{fig:2}) we cannot neglect the delayed terms $v_\tau\equiv v(t-\tau)$ and $u_\tau\equiv u(t-\tau)$ in Eq.~(\ref{eqn:gen1}):
\begin{equation}
\begin{aligned}
\label{eq:}
\varepsilon \dot{u} &= u - \frac{u^3}{3} - v + \sigma (v_\tau-v)\\
\dot{v} &= u + a - \sigma (u_\tau-u)
\end{aligned}
\end{equation}
Due to the almost linear behavior on the slow branches (exemplarily shown by the straight connection between $u(t)$ and $u(t-\tau)$ in Fig.\,\ref{fig:3}a) we assume $\vec{x}(t)-\vec{x}(t-\tau)=\tau \dot{\vec{x}}(t)$ for values of $\tau$ close to multiples of the period $mT$ with $m \in \mathbb{N}$:
\begin{equation}
\begin{aligned}
\label{eq:}
\varepsilon \dot{u} &= u - \frac{u^3}{3} - v - \sigma \tau \dot{v}\\
\dot{v} &= u + a + \sigma \tau \dot{u}
\end{aligned}
\end{equation}
We can insert the second equation into the first one and analyze the dynamics on the falling branches of the $u$-nullcline given by $v = u - \frac{u^3}{3}-\sigma \tau (u+a)$:
\begin{eqnarray}
\dot{u}=\frac{u+a}{1 -u^2-2\sigma\tau}.
\end{eqnarray}
This is an approximation of the equation which would have been obtained if the phase lag term $\cos \phi$ were not been neglected:
\begin{eqnarray}
\label{eq:T_phaselag}
\dot{u}=\frac{\frac{u+a}{1+\sigma \tau \cos \phi}}{1-u^2-2\sigma\tau\tfrac{ \sin \phi}{1+\sigma \tau \cos \phi}}.
\end{eqnarray}
In the case of values of $\tau$ close to $T$ we can calculate the period of the synchronized oscillations as
\begin{eqnarray}
T_{sync}(\tau) \propto u_+^2-u_-^2+\left(1-a^2-2\sigma \tau\right) \ln \frac{a^2-u_-^2}{a^2-u_+^2}
\label{eq:T_tau}
\end{eqnarray}
As proportionality factor in Eqs.\,\eqref{eq:T_sigma} and \eqref{eq:T_tau} we assume $1+e(\varepsilon)$, where $e(0.05)=0.3$ is a constant parameter, determined by fitting the analytical solution (Eq.\,\eqref{eq:T_sigma} for $\sigma=0$) to the numerical simulation (Eq.\,\eqref{eqn:gen1} for $\sigma=0$). As generally shown in \cite{YAN09}, delay systems generically have branches of periodic solutions, which are reappearing for integer multiples of the intrinsic period $T$ of the system. A solution for $\tau = \tau_0 < T$ reappears for all values
\begin{eqnarray}
\label{eq:tau_0}
\tau_m = \tau_0 + mT_{sync}(\tau_0)
\end{eqnarray}
with $m \in \mathbb{N}$, and $T_{sync}$ depends upon $\tau_0$ according to Eq.\,\eqref{eq:T_tau}. The branches $T_{sync}$ of the synchronized solutions are piecewise linear functions of $\tau$, as shown in Fig.\,\ref{fig:3}c, where $m = 1,2,...$ numbers the branches. With increasing $m$ the branches are stretched by $\partial \tau_m \over \partial \tau_0$ and their slope decreases (see \cite{YAN09}). To take into account this mapping for $m>0$, $\tau$ in Eq.\,\eqref{eq:T_tau} has to be replaced by 
\begin{eqnarray}
\tau' =  \tau_0  \left(\partial \tau_m \over \partial \tau_0 \right)^{-1} =  \tau_0 \left(1-2\sigma m (1+e) \ln \frac{a^2-u_-^2}{a^2-u_+^2}\right)^{-1},
\label{eq:tau_m}
\end{eqnarray}
where for a given $\tau = \tau_m>T$, $\tau_0$ and $m$ can be calculated from Eq.\,\eqref{eq:tau_0}. Eq.\,\eqref{eq:T_tau} now reads 
\begin{eqnarray}
T_{sync}(\tau') = (1+e)\left[u_+^2-u_-^2+\left(1-a^2-2\sigma \tau'\right) \ln \frac{a^2-u_-^2}{a^2-u_+^2}\right].
\label{eq:T_tau_gen}
\end{eqnarray}
A comparison of this analytical result for the period $T_{sync}$ in the synchronized regime with numerical simulations is given in Fig.\,\ref{fig:3}c. Depending on the initial conditions, we can find chimera states in the red shaded regions at the boundaries of the piecewise linear branches, which occur if the delay times $\tau$ are half-integer multiples of the intrinsic period $T_{sync}(0) = T$.
They are marked in Fig.\,\ref{fig:3}c (red shaded) for $\sigma=0.15$. Note that the period is piecewise linear as a function of $\tau$ and also of $\sigma$ (in case of $m=0$ in Eq.\,\eqref{eq:tau_m}) in the synchronized regime, whereas it is nonlinear in the non-synchronized regime.

In addition we can see a decrease of the maximum of the chimera tongues with increasing $\tau$ in Fig.\,\ref{fig:2}, cf.~\cite{JUS97,JUS00,JUS04}: The maximal value of the coupling strength for which chimera states can be observed decreases for increasing delay.

\begin{figure}[htbp]
\centering
\includegraphics[width = 1.01\textwidth]{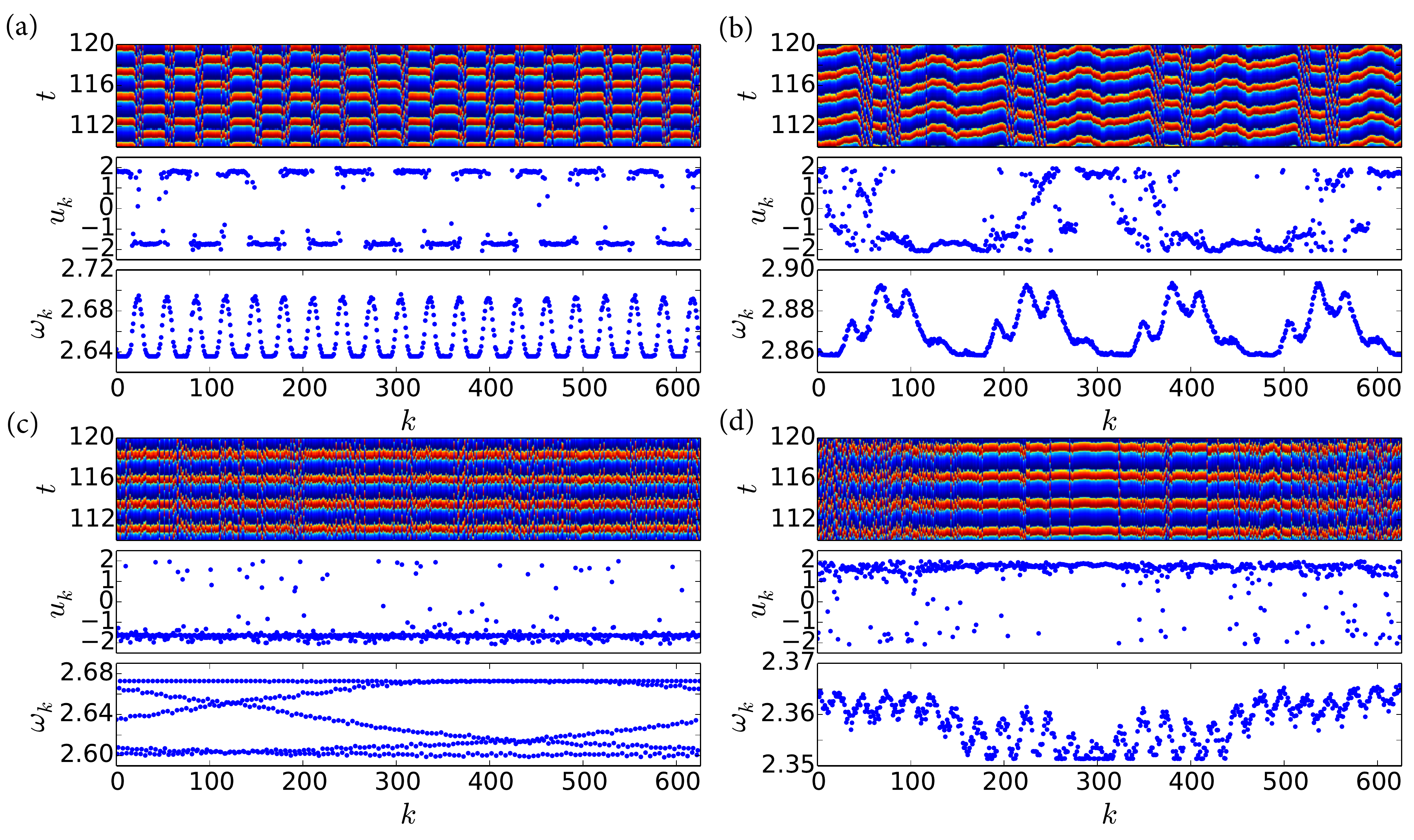}
\caption{(Color online) Patterns occurring in the chimera tongues in Fig.\,\ref{fig:2}: Space-time plot of $u$ (upper panels), snapshot of variables $u_k$ (middle panels), and mean phase velocity profile $\omega_k$ (bottom panels) for (a) $\tau = 1.0$ and $\sigma=0.1$, (b) $\tau = 1.0$ and $\sigma=0.15$, (c) $\tau = 1.4$ and $\sigma=0.05$, and (d) $\tau = 10.7$ and $\sigma=0.01$. Other parameters as in Fig.\,\ref{fig:1}.}
\label{fig:4}
\end{figure}

Let us now take a closer look at the dynamics inside the tongues in Fig.\,\ref{fig:2}. For the parameter values chosen inside the first, leftmost and largest, tongue we find multichimera states (which consist of several coherent and incoherent parts, here $20$ each, i.e., we have a $20$-chimera) similar to Fig.\,\ref{fig:1} (see Fig.\,\ref{fig:4}a) and nested chimera structures (see Fig.\,\ref{fig:4}b). These nested structures are slowly shifting in space, so that the mean phase velocity profile (bottom panel of Fig.\,\ref{fig:4}b) shows a pyramidal structure instead of an arc-like profile as usually in stationary chimera states. The speed of traveling is sensitive to the coupling strength and delay time. For a pronounced profile of the mean phase velocity this speed should be small. Otherwise it is smeared out over time. Fig.\,\ref{fig:4}c and d show two examples of the transition region from complete synchronization to chimera states. Also here we have coherent and incoherent domains. In contrast to the other examples we can find a complex structure of the mean phase velocity profiles (see bottom panels). In general, the appropriate choice of time delay $\tau$ in the system allows one to achieve the desired chimera pattern.  

\begin{figure}[htbp]
\centering
\includegraphics[width = 1.01\textwidth]{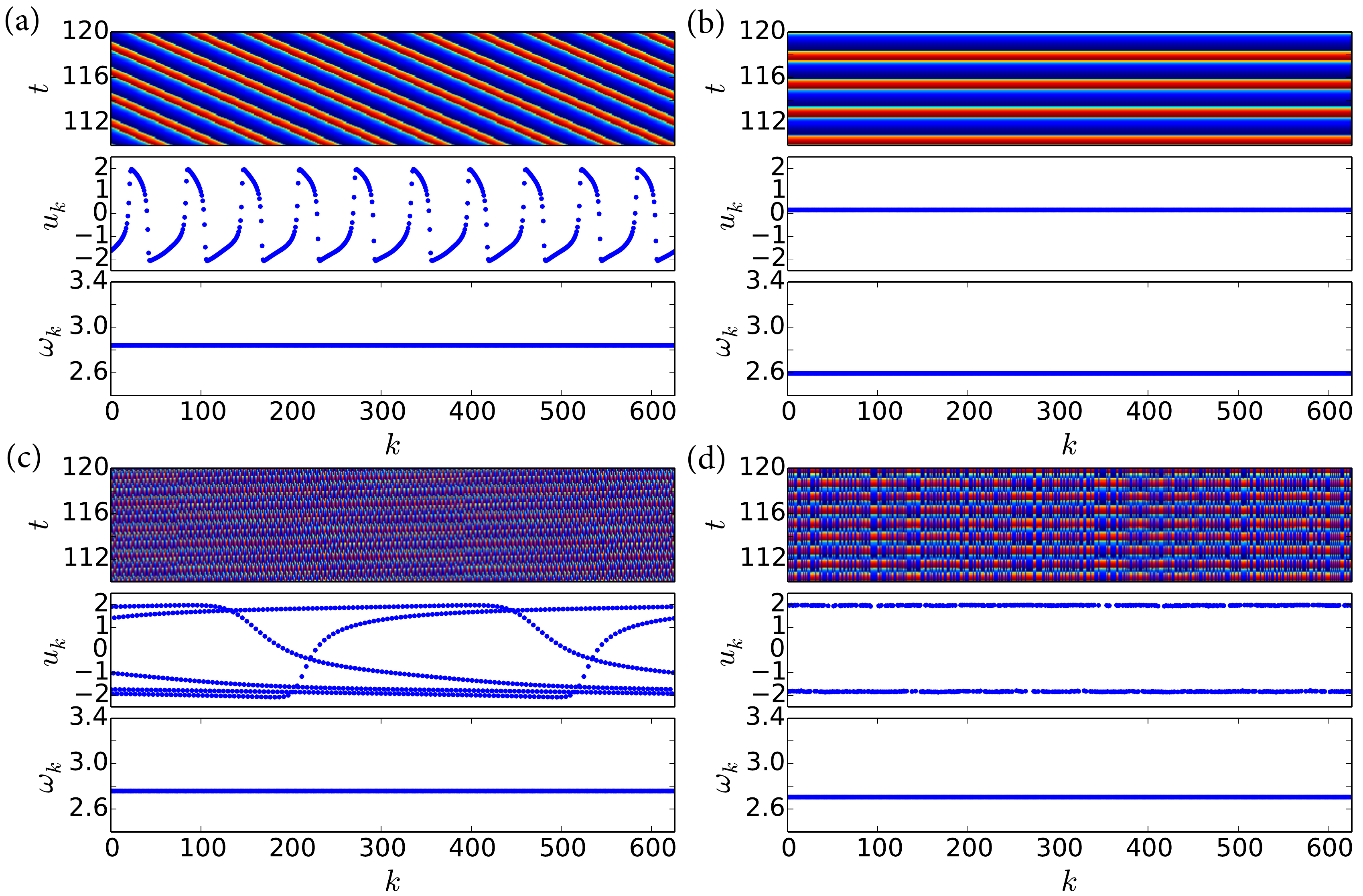}
\caption{(Color online) Patterns occurring in non-chimera regimes of Fig.\,\ref{fig:2}: Space-time plot of $u$ (upper panels), snapshot of variables $u_k$ (middle panels), and mean phase velocity profile $\omega_k$ (bottom panels) for (a) $\tau = 1.1$ and $\sigma=0.15$, (b) $\tau = 4.5$ and $\sigma=0.1$, (c) $\tau = 5.1$ and $\sigma=0.1$, and (d) $\tau = 5.5$ and $\sigma=0.1$. Other parameters as in Fig.\,\ref{fig:1}.
}
\label{fig:5}
\end{figure}

In the parameter plane of delay time $\tau$ and coupling strength $\sigma$ the region corresponding to coherent states is dominating (blue and yellow regions in Fig.\,\ref{fig:2}). On one hand, we observe the in-phase synchronization regime (see Fig.\,\ref{fig:5}b) which is enlarged for increasing coupling strength. On the other hand, we also detect a region of coherent traveling waves with wavenumber $k>1$ (see Fig.\,\ref{fig:5}a) and $k < 1$ (see Fig.\,\ref{fig:5}c). Varying the delay time~$\tau$ allows not only for switching between these states, but also for controlling the speed of traveling waves: in the diagonal striped yellow region in Fig.\,\ref{fig:2} the mean phase velocity decreases for increasing delay times. In addition we can observe {\em salt and pepper} states (see Fig.\,\ref{fig:5}d), where all nodes oscillate with the same phase velocity but they are distributed between states with phase lag $\pi$ incoherently~\cite{BAC14}. As discussed above, the reason for this are arbitrarily short wavelengths of the dynamical patterns.

\section{Discussion}

In the current study, we have analyzed chimera states in ring networks of FitzHugh-Nagumo oscillators with hierarchical connectivities. For a fixed base pattern, we have constructed a hierarchical connectivity matrix, and provided a numerical study of complex spatio-temporal patterns in the network. Our study was focused on the role of time delay in the coupling term and its influence on the chimera states.

In the parameter plane of time delay~$\tau$ and coupling strength~$\sigma$, we have determined the regimes for different types of chimera states, alternating with regimes of coherent states. An appropriate choice of time delay allows us to stabilize several types of chimera states. The interplay of complex hierarchical network topology and time delay results in a plethora of patterns going beyond regular two-population or nonlocally coupled ring networks: we observe chimera states with coherent and incoherent domains of non-identical sizes and non-equidistantly distributed in space. Moreover, traveling and non-traveling chimera states can be obtained for a proper choice of time delay. We also demonstrate that time delay can induce patterns which are not observed in the undelayed case. In addition we have shown analytically the influence of $\tau$ upon the period; i.e., the phase velocity, a piecewise linear dependence in regimes with coherent states, whereas a nonlinear dependence upon $\tau$ is found for incoherent states.

Our analysis has shown that networks with complex hierarchical topologies, as arising in neuroscience, can exhibit diverse nontrivial patterns. Time delay can play the role of a powerful control mechanism which allows either to promote or to destroy chimera patterns.

\begin{acknowledgement}
This work was supported by DFG in the framework of SFB 910. We are grateful to Serhiy Yanchuk, Rico Berner, and Denis Nikitin for insightful discussions.
\end{acknowledgement}

\section{Authors contributions}
JS did the numerical simulations and the theoretical analysis. IO, AZ and ES supervised the study. All authors designed the study and contributed to the preparation of the manuscript. All the authors have read and approved the final manuscript.

\bibliography{ref}
\bibliographystyle{prsty-fullauthor}

\end{document}